\documentclass[12pt]{article}
\usepackage{epsfig}
\usepackage{cite}
\def\dfrac#1#2{{\displaystyle{#1\over#2}}}
 
\begin{document}
 \begin{flushright}
        DESY 01-117 \\ 
        August 22, 2001
\end{flushright}\bigskip\bigskip\bigskip\bigskip\bigskip
 \begin{center}
   {\huge\bf 
     Electromagnetic Radiation \\ \vspace{1mm} Hardness of Diamond Detectors
   }
 \end{center}\bigskip\bigskip
 \begin{center}{{\large 
       T.~Behnke~$^a$, M.~Doucet~$^a$\footnote{Corresponding author}, 
       N.~Ghodbane~$^a$, \\ A.~Imhof~$^b$, C.~Mart\'{\i}nez~$^c$, 
       W.~Zeuner~$^a$\\
               }
\bigskip
     {\normalsize 
       $^a$ DESY, Hamburg, Germany \\
       $^b$ Universit\"at W\"urzburg, W\"urzburg, Germany \\
       $^c$ University of Portsmouth, Portsmouth, UK  }
     }\end{center}\bigskip\bigskip
 \bigskip\begin{center}{\large  Abstract}\end{center}
\begin{center}\begin{minipage}{5.2truein}
The behavior of artificially grown CVD diamond films under intense
electromagnetic radiation has been studied.  The properties of 
irradiated diamond samples have been investigated using the method
of thermally stimulated current and by studying their charge 
collection properties.  Diamonds have been found to remain unaffected 
after doses of 6.8~MGy of 10~keV photons and 10~MGy of MeV-range photons.
This observation makes diamond an attractive detector material for
a calorimeter in the very forward region of the proposed TESLA detector.
\end{minipage}\end{center}
\bigskip\bigskip\bigskip\bigskip
\bigskip\bigskip
\begin{center}
{\large (To be submitted to Nucl. Instr. and Meth.)}
\end{center}

\section{Introduction}
\label{sect-intro}
Diamond has been extensively studied in recent years for use for particle 
detection~\cite{adam1}.  Many studies have focused on the hadronic radiation 
hardness properties of diamond for applications at the LHC.  Diamond detectors 
have been shown to remain undamaged up to fluences of neutrons, protons and pions
of the order of $10^{15}$/cm$^2$~\cite{adam2}.  On the other 
hand, diamonds have only been tested with electromagnetic doses of up to 
0.1~MGy~\cite{bauer}.  Electromagnetic radiation hardness is a very important 
issue for low-angle detectors for the TESLA linear accelerator project~\cite{tdr}.
A luminosity calorimeter, which would serve both as a fast luminosity
monitor~\cite{napoly} and as a low-angle calorimeter, is planned for this project.
According to the current design, this detector would be placed only 1.2~cm
from the beam line.  A very large beam-induced background of photons and 
electrons would deposit a dose of as much as 1~MGy per year in the detector 
elements closest to the beam.  Given these very large doses and the resistance 
of diamond to hadronic radiation, diamond has been considered as a potential 
detector material.  Silicon detectors have also been shown to operate after 
having received doses of several MGy of electromagnetic radiation~\cite{si} 
and could also be considered as active material for this detector.  This paper 
addresses the electromagnetic radiation hardness properties of diamond up to 
doses comparable to several years of TESLA running.  Collection distance 
measurements and thermally stimulated current measurements (TSC) were conducted 
using diamonds irradiated by a 10~keV photon beam.  Collection distance 
measurements were also performed with a diamond irradiated by a $^{60}$Co source 
(emitting $\beta^{\pm}$ with 0.32~MeV endpoint and photons of 1.17~MeV and 1.33~MeV).

\section{Principle of operation}
Artificially grown CVD diamond has already been demonstrated to be a good 
semi-conductor detector material~\cite{adam1}.  A voltage is applied between 
two electrical contacts made on each side of a thin diamond film, typically 
of the order of 300-500~$\mu$m thick.  The energy deposited by passing charged 
particles creates electron-hole pairs which induce a charge on the contacts as 
they migrate towards them.  The band-gap of diamond is 5.5~eV, and 13~eV is on 
average needed to create an electron-hole pair.  

Due to impurities in the diamond, the migrating charges can be trapped on their 
way towards the electrical contacts.  The charge induced on the contacts is then 
smaller than the total charge created in the diamond.  The collection efficiency 
$\epsilon$, is defined as 
\begin{equation}
  \epsilon = \frac{ Q_{{\rm induced}} } { Q_{{\rm deposit}} }, 
\end{equation}
where $Q_{{\rm induced}}$ is the charge induced on the contacts and 
$Q_{{\rm deposit}}$ corresponds to the charge of the total number of electrons
released in the diamond by the ionization process.  The collection distance, 
representing the average distance between the electron and the hole of a given 
pair at the moment when they are stopped, is related to the collection efficiency by 
\begin{equation}
  \delta = \epsilon D,
\end{equation}
where $D$ is the thickness of the detector~\cite{adam1}.  For CVD diamonds, the 
collection distance is usually significantly shorter than the thickness of the 
detector (up to about 2/3 of its thickness).  Being intimately linked to the 
performance of the detector, the collection distance is used as a figure of merit 
of the diamond material.

The collection distance is influenced by the applied voltage.  Since the mobility 
of charge carriers depends on the applied voltage, the collection distance will in 
turn depend on the voltage.  It reaches an asymptotic value at a voltage 
corresponding to the saturation point of the mobility.  In the following, we will 
always refer to the asymptotic value of the collection distance, reached at a 
typical voltage of 1 V/$\mu$m.

The collection distance is affected by the amount of radiation the detector has 
previously received, through the so-called priming effect~\cite{oh}.  As the diamond 
is irradiated, the charges created in the material are partially trapped in energy 
levels that are created by impurities.  Although short-lived defects 
can be expected to rapidly release the trapped charges and be available for trapping 
at a later time, long-lived defects can remain filled long enough to be unavailable 
for further trapping on the time-scale of the measurement.  Such long-lived defects 
have been observed in diamond using the method of thermally stimulated 
current~\cite{bowlt, gonon, tromson, briand}.  The priming effect decreases the 
effective number of trap energy levels and the collection distance consequently 
increases.  This effect reaches saturation once all long-lived traps have been filled.  
In the following, we will always refer to the collection distance of primed diamonds.

\section{Thermally stimulated current measurements}
\label{sect-tsc}
The energy and the amount of trap levels in the diamond can be measured with the 
method of thermally stimulated current~\cite{tsc}.  A voltage is applied between 
contacts on both sides of the diamond.  One of the two contacts is grounded and 
is used to measure the current flowing through the material.  After a short 
irradiation period to fill traps, the diamond is heated up.  As the temperature 
of the diamond rises, trapped charges are thermodynamically released at a rate 
depending on the temperature and on the energy level of the traps.  A current 
proportional to the trap density and to the release rate is then observed between 
the contacts of the diamond sample.  As the temperature rises, the release rate 
increases until no charge is left to be released.  For large enough temperature, 
the rate of electrons in the valence band of the diamond thermodynamically 
crossing the band-gap can be sufficient to create a large current.  This current 
is increasing with the temperature and constitutes a background to the TSC from 
released traps.

The TSC dependence on the temperature gives information on the energy levels and 
the density of the impurities in the diamond.  Since the measured current 
corresponds to a rate of release, the heating rate of the diamond also has to be 
known to extract this information.  

In the present TSC measurements, a sizeable portion of the TSC comes from the 
background current at large temperature.  This translates into a large error on 
the background subtraction.  
For this reason, only the TSC peak maximum, for which the background contribution 
is smaller than at higher temperature, has been used for comparing the diamond 
behavior after various doses of radiation.  A fit to the TSC curves has also 
been done to extract the energy level of the TSC peak and verify the quality
of the data.

Assuming that released traps have a negligible probability of being recaptured, 
the TSC can be written as follows~\cite{Mckeever}:
\begin{eqnarray}
I(T) = s n_0  \exp \left( -\dfrac{E}{k_B T}\right) 
              \times \exp \left[
              -\dfrac{s}{\beta}
		 \int_{T_0}^{T}\mbox{d}T \exp\left(-\dfrac{E}{k_B T}\right)
		\right],
\label{eqn1}
\end{eqnarray}
where $I$ is the TSC, $n_0$ is the initial number of trapped carriers, $E$ the 
energy level of the trap, $s$ a frequency factor, $T_0$ the initial temperature, 
$T$ the absolute temperature, $k_B$ the Boltzmann constant and $\beta$ the heating 
rate.

In this paper, the energy levels were extracted using an approximation~\cite{kitis}
of Equation~\ref{eqn1}:
\begin{eqnarray}
I(T) = I_m       \exp \left[ 1 + \dfrac{E}{k_B T} \dfrac{T-T_m}{T_m} 
		               - \dfrac{T^2}{T_m^2} 
\left( 1 -  \dfrac{2 k_B T}{E}\right)
				\exp\left(\dfrac{E}{k_B T} \dfrac{T-T_m}{T_m}\right)
			      - \dfrac{2 k_B T_m}{E}
		\right],
\label{eqn2}
\end{eqnarray}
where $I_m$ and $T_m$ are the intensity and the absolute temperature at the 
peak maximum, which are related to $\beta$, $s$ and $n_0$ by
\begin{eqnarray}
I_m = n_0 \dfrac{\beta E}{k_B T_m^2 }\exp\left(\dfrac{2 k_B T_m}{E} 
- 1 \right) \mbox{ and } s =  \dfrac{\beta E}
{k_B T_m^2 }\exp\left(\dfrac{E}{ k_B T_m}\right).
\label{eqn3}
\end{eqnarray}

\section{Experimental procedures}
The Hasylab facility at DESY provides photon beams from synchrotron
radiation.  A 10~keV photon beam~\cite{hasylab} was used to expose 
two diamond samples to large electromagnetic radiation doses.  The 
average irradiation rate was calculated to be approximately equal to 
14~Gy/s.  To test the resistance of the diamond samples to radiation, 
two types of measurements were performed before and after the 
irradiation periods.  The collection distance of each sample was 
measured and a set of TSC measurements was performed.  During the TSC 
measurements, the Hasylab beam was also used to create electron-hole 
pairs in the sample under study.  To verify the resistance to radiation 
from photons having a higher energy, a $^{60}$Co source was also used 
to irradiate a third diamond sample.

The collection distance was evaluated by measuring the total charge
induced on the contacts of the sample by electrons from a $^{90}$Sr 
source.  A signal from a Si-detector placed behind the diamond detector 
provided a trigger.  For each trigger, the diamond detector was read out 
using an Amptek~A250~\footnote{AmpTek Electronics} pre-amplifier, that was 
followed by an Ortec amplifier/shaper with 3~$\mu$s shaping time.  Both 
the Si-detector and diamond signals were read out by a digital oscilloscope.  
The collection distance is related to the measured signal left by single 
electrons by the following relation:
\begin{equation}
  \delta = \frac{ Q_{{\rm meas}} } { Q_{{\rm deposit}} } D,
\end{equation}
where $Q_{{\rm meas}}$ is the total measured charge, $Q_{{\rm deposit}}$
corresponds to the predicted charge of the total number of electrons 
released by the ionization process of a single incident electron and $D$ 
is the thickness of the diamond.  The measurements were carried out after 
having primed the diamonds, at a voltage corresponding to the maximum 
carrier velocity.

To measure the TSC, a nominal voltage of -50~V was applied on one contact 
of the diamond during measurements and irradiation periods.  The other contact 
was connected to a Keithley~6514~\footnote{Keithley Instruments GmbH} to 
measure the current.  To generate the thermally stimulated current, a 
remote-controlled heating element was used.  The heating rate of this element
was measured to be $4.6 \pm 0.3$ K/s.  The temperature was monitored using a 
thermocouple element read by a voltmeter.  During the acquisition sequence, 
the temperature and the current were measured at two second intervals.  To 
reduce the time between TSC measurements, liquid nitrogen was used to cool the 
heating element to ambient temperature once the maximum temperature was reached.

\begin{figure}
  \begin{center}
    \mbox{\epsfig{file=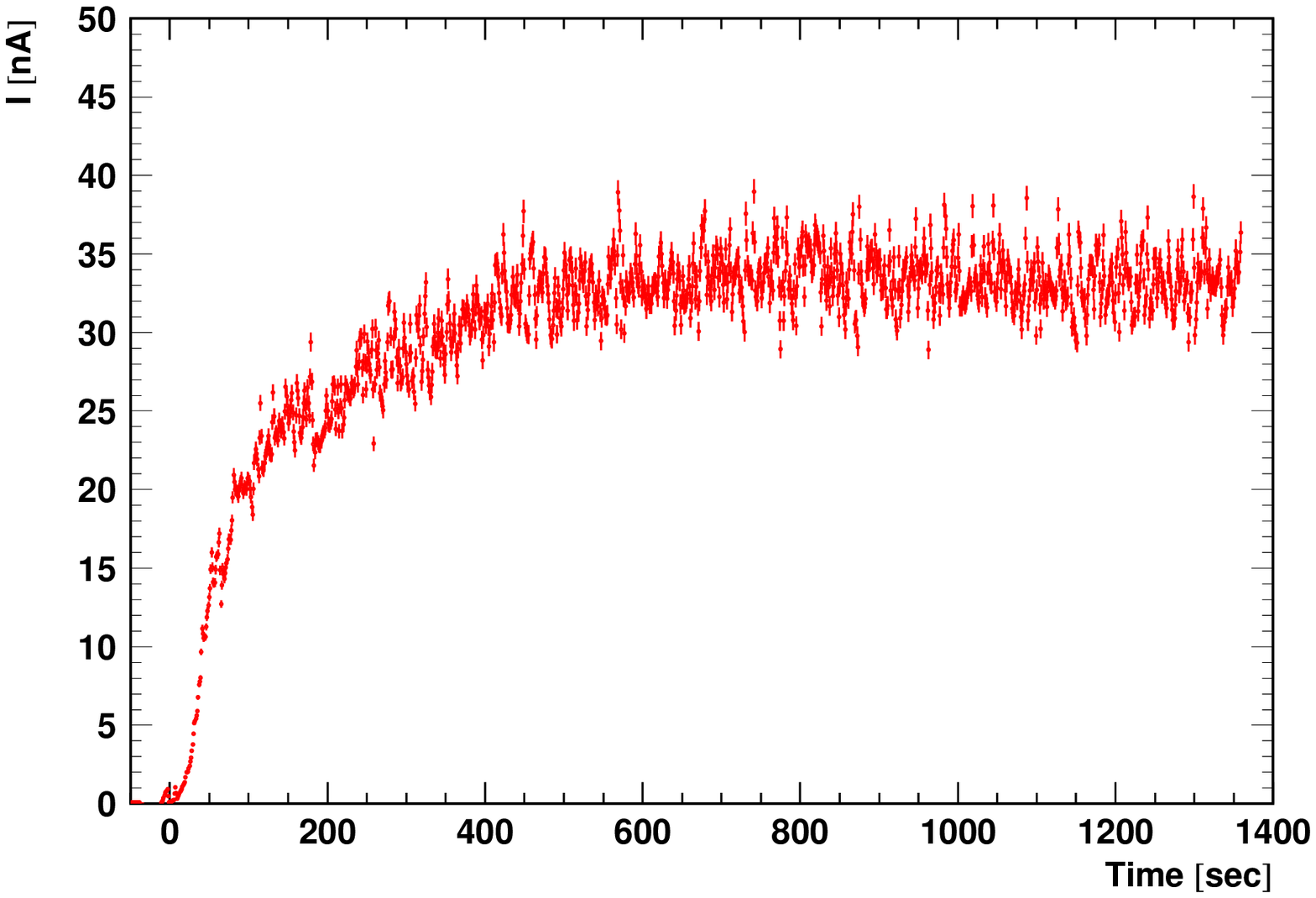,width=16cm}}
    \caption{\label{fig-tirr}
     Current as a function of time during irradiation.
      } 
  \end{center}
\end{figure}

Before heating the diamond, a fixed period of irradiation was used to create 
electron-hole pairs in the diamond and fill traps.  For this purpose, the 
10~keV photon beam of Hasylab was directed on a $100\times 100 \mu{\rm m}^2$ 
slit.  A remote-controlled shutter was placed between the detector and the 
slit to switch the beam on and off between data taking periods.  To monitor 
the photon beam, a scintillator read out by a photomultiplier was placed near 
the beam to measure the electrons produced by the scattering of beam photons 
with air.  For each acquisition sequence, an irradiation period of 60~seconds 
was done before the TSC curve was recorded, corresponding to approximately 840~Gy.  
Figure~\ref{fig-tirr} shows the current measured on one contact of the diamond 
during an irradiation period of more than 20~minutes.  The current rises as 
traps in the diamond are filled and reaches a maximum after about 10~minutes 
($\sim 8.4$~kGy).  The 60~second irradiation time used is therefore well below 
the time period over which saturation effects are visible.

Three diamonds of similar properties were used for three sets of measurements.  
All three were produced by DeBeers~\footnote{DeBeers Industrial Diamond Division}.  
They have a thickness of 300~$\mu$m and an average grain size of approximately 
20~$\mu$m.  Contacts were deposited on both sides of each 
sample~\footnote{The contacts were done by the Fraunhofer Institut f\"ur Schicht 
und Oberfl\"achentechnik, Braunschweig, Germany}.  They consist of depositions 
of Ti~(50~nm), Pt~(30~nm) and Au~(60~nm).

One diamond (denoted diamond~\#1) was used to perform a surface scan of TSC 
measurements.  A region of about 1~mm$^2$ in steps of $100\times 100 \mu{\rm m}^2$ 
was measured before irradiation and after doses of about 0.2~MGy and 1.4~MGy.

A second diamond (\#2) was used in a test involving larger doses.  TSC 
measurements were performed at various voltage values before irradiation and 
after doses of about 1.4~MGy and 5.4~MGy.  For these measurements, the full 
surface of the diamond was irradiated.  For both diamonds (\#1 and \#2), the 
collection distance was measured before irradiation and after the last irradiation.

A third diamond (\#3) was sent for irradiation at a $^{60}$Co 
source~\footnote{Gamma-Service Produktbestrahlung GmbH}.  The collection 
distance of the diamond was also measured before irradiation and after doses
of 1~MGy and 10~MGy.  A $^{60}$Co emits photons of 1.17~MeV and 1.33~MeV, 
which are above the threshold for non-ionizing damages.  It is therefore 
interesting to compare the effects of low energy and high energy photon 
radiation.

\section{TSC signal treatment}
\label{sect-signal}
\begin{figure}
  \begin{center}
    \mbox{\epsfig{file=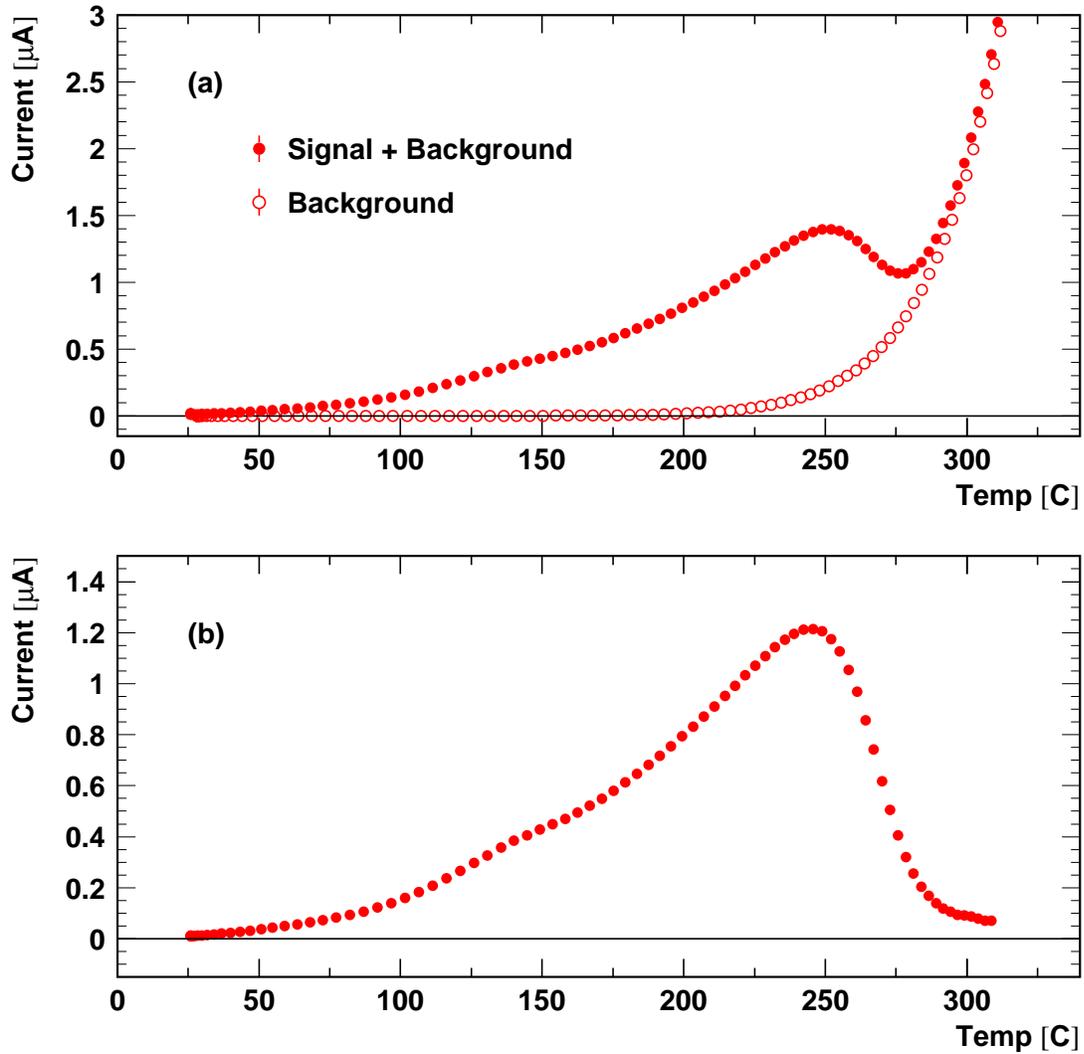,width=16cm}}
    \caption{\label{fig-ex}
    Example of TSC signal and background subtraction.
    Figure~(a) shows the TSC signal measured after 60~seconds of
    irradiation~(points) and a background signal taken
    in another run~(circles).  Figure~(b) shows the
    signal after background subtraction.
      } 
  \end{center}
\end{figure}

Figure~\ref{fig-ex}a shows an example of a TSC signal measured with
diamond~\#1 during the surface scan.  The peak at 250~$^{\circ}$C corresponds 
to the TSC peak from released charges.  The background current described in 
Section~\ref{sect-tsc} can be seen at higher temperature and  must be 
subtracted to obtain the part of the current due to the trap release.  The 
open points on Figure~\ref{fig-ex}a show a background curve obtained in a 
subsequent run without irradiation.  For each TSC curve of the scan performed
with diamond~\#1, the signal was extracted by subtracting the same background 
curve.  Figure~\ref{fig-ex}b shows the signal extracted from Figure~\ref{fig-ex}a.

In order to verify the background subtraction technique and to estimate the 
systematic errors, each scan was subdivided in four small raster scans of
$0.5\times 0.5 \mu{\rm m}^2$ with overlapping regions.  The overlapping points, 
taken at different times, were compared and the RMS of the distribution of the 
difference between the measurements was taken as the systematic error.  A systematic 
error of 12\% was found.  This error includes the effect of the flux variations, 
for which no correction was applied for this evaluation.

For the TSC data taken with diamond~\#2 (for which the complete surface of the 
diamond was irradiated), the background subtraction scheme was different.  Due 
to the large irradiated surface, the TSC signals were very large.  The background 
was changing according to the voltage applied, which was varied between 5~V and 
150~V.  A fifth degree polynomial was fitted on each side of the TSC peak to 
estimate the background value at the peak maximum.  The results obtained with 
this technique were checked by comparing to background curves taken at 50~V.  A
conservative error equal to the fitted background current, corresponding to 
about 8\% of the TSC peak, was added to the 12\% systematic uncertainty mentioned 
earlier.  The total error on the TSC peak is taken to be 15\%.

\begin{figure}
  \begin{center}
    \mbox{\epsfig{file=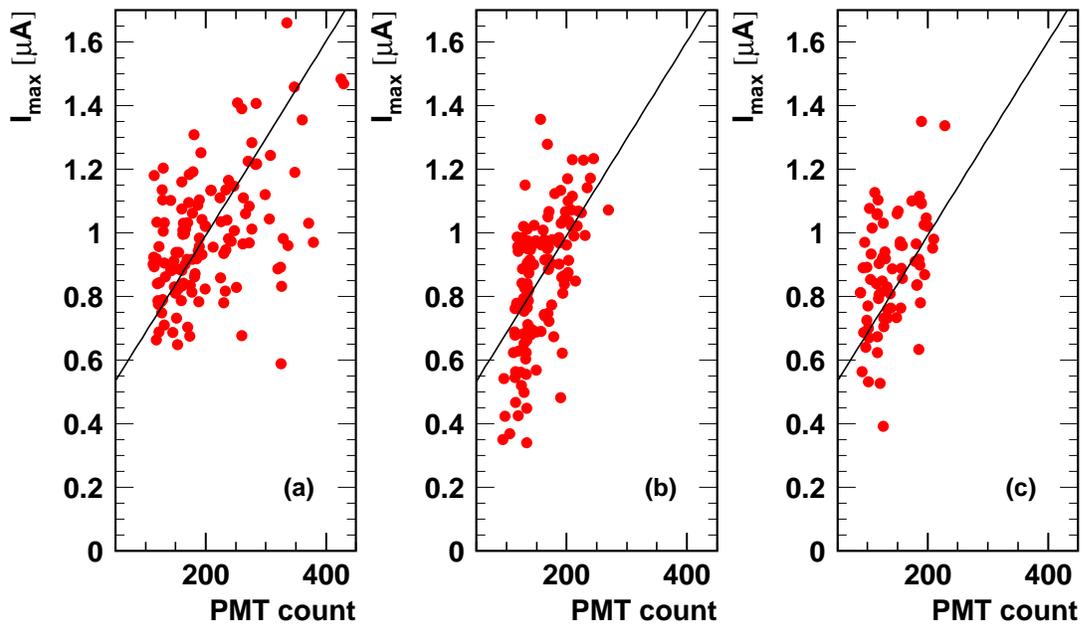,width=16cm}}
    \caption{\label{fig-flux}
    Correlation between beam flux (PMT counts) and TSC,
    before irradiation~(a), after 0.2~MGy~(b) and after 1.6~MGy~(c).
    Note that the beam flux corresponding to the measurements done
    before irradiation~(a) reached values much larger than in the
    other two cases.
      } 
  \end{center}
\end{figure}

Although the background remains the same for all TSC curves taken at a given 
voltage, the signal depends on the photon beam flux.  In order to apply a 
correction for the beam flux, a scintillation counter was placed near the beam 
to count beam-air interactions.  Figure~\ref{fig-flux} shows the correlation 
between the maximum TSC and the beam flux for the three scans of diamond~\#1.
If no flux variation were present, no correlation between the two should be 
visible.  A linear fit to the points is also shown.  A linear flux dependence 
of the TSC signal is justified by the short irradiation period of 60~seconds.
In such a short time, no saturation effect of the trap filling is expected
(see Figure~\ref{fig-tirr}).  This linear relation was used to correct the TSC 
signal to a single photon flux value.  The systematic error corresponding to 
the flux correction is included in the 12\% uncertainty mentioned above.  

\section{Extraction of energy levels and trap density}
\label{sect-levels}
A fit to the TSC curves was performed to evaluate the energy level of the traps.  
The large sample of curves from the TSC scan was also used to evaluate the 
systematic error on this measurement.  The comparison of the results to 
previously published data~\cite{bowlt, gonon, tromson, briand} provides a 
cross-check of the quality of the TSC measurements.

\begin{figure}
   \begin{center}
      \begin{tabular}{cc}
         \includegraphics[width=8cm,height=8.5cm]{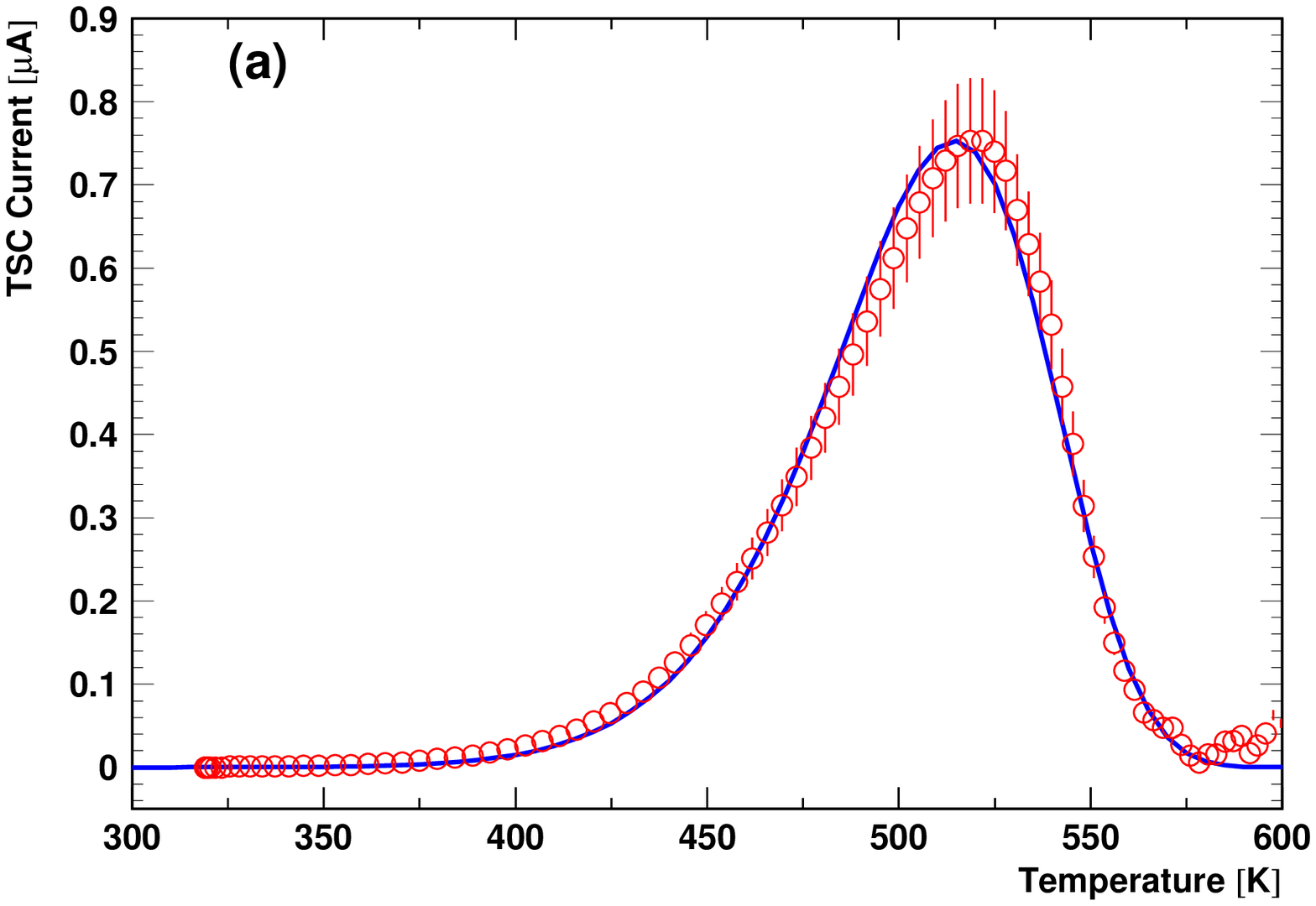}&
         \includegraphics[width=8cm,height=8.5cm]{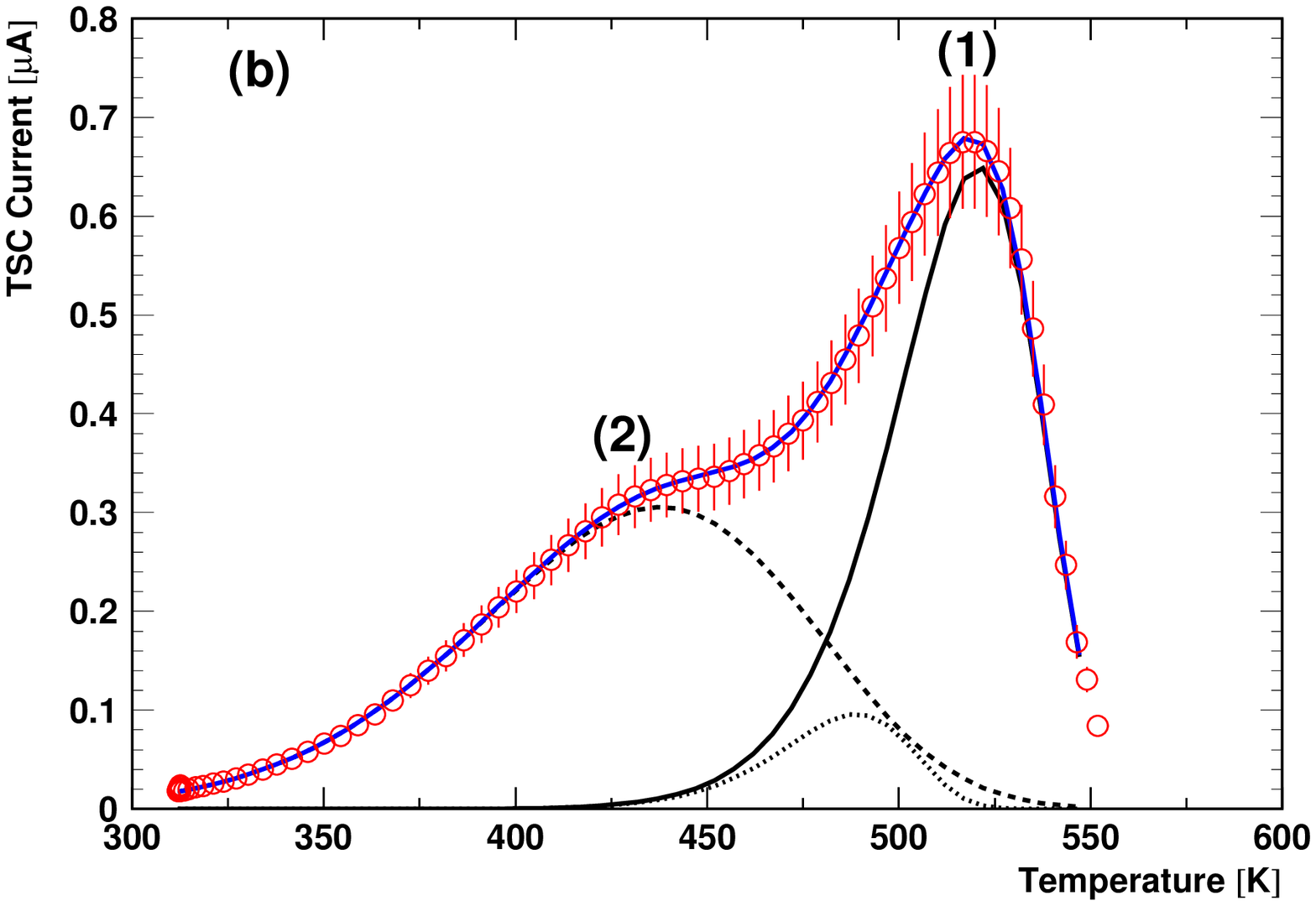}\\
      \end{tabular}
      \caption{\label{fig:r46e23.eps} 
  Fit of the data to equation~(\ref{eqn2}).  
  Figure~(a) corresponds to an energy level of $E=0.83 \pm 0.06 \pm 0.24$~eV and 
  Figure~(b) shows two peaks at $E_1 = 1.186 \pm 0.002 \pm 0.360$~eV and 
  $E_2 = 0.3456 \pm 0.0003 \pm 0.1000$~eV.}
   \end{center}
\end{figure}

Figures~\ref{fig:r46e23.eps}a and~\ref{fig:r46e23.eps}b show two examples of 
TSC curves and the associated fits using Equation~\ref{eqn2}.  The two curves 
correspond to diamond~\#1, with an irradiated area of 
$100\mu{\rm m} \times 100\mu{\rm m}$.  The data shown on 
Figure~\ref{fig:r46e23.eps}a were fitted using a single TSC peak, while a 
superimposition of three TSC peaks was fitted to the data shown on 
Figure~\ref{fig:r46e23.eps}b.  For Figure~\ref{fig:r46e23.eps}a, 
an energy level equal to 
$E=0.83 \pm 0.06({\rm stat}) \pm 0.24({\rm syst})$~eV was found.  
The systematic error on all fitted parameters was estimated using the discrepancy 
between several measurements.  The quality of the fit is estimated with the 
figure of merit (FOM) of Balian and Eddy~\cite{balian}, in the present case 
equal to $0.026$.  Using Equation~(\ref{eqn3}), one gets a frequency factor 
$s = 2.38 \pm 0.01({\rm stat}) \pm 1.2({\rm syst})$~GHz.  On 
Figure~\ref{fig:r46e23.eps}b, one can clearly notice the presence of 
two trap energy levels fitted at 
$E_1 = 1.186 \pm 0.002({\rm stat}) \pm 0.360({\rm syst})$~eV for the most 
intense and $E_2 = 0.3456 \pm 0.0003({\rm stat}) \pm 0.1000({\rm syst})$~eV 
for the second one.  The FOM for this fit is equal to $0.00013$.  Using all 
TSC curves from diamond~\#1, an average energy of 
$E_1 = 1.10 \pm 0.03({\rm stat}) \pm 0.30({\rm syst})$~eV was measured for the peak
at 250~$^{\circ}$C and 
$E_2 = 0.57 \pm 0.04({\rm stat}) \pm 0.17({\rm syst})$~eV was measured for the peak 
at around 150~$^{\circ}$C.

Different energy levels of defects found in CVD diamond samples are reported 
in the literature.  Two energy ranges have been observed, one below 1~eV 
(0.5~eV~\cite{bowlt}, 0.87~eV~\cite{tromson} and 0.8~eV~\cite{briand}) and one 
above 1~eV (1.8~eV~\cite{bowlt}, 1.86~eV~\cite{gonon} and 1.42~eV~\cite{tromson}).  
Our results are compatible with the low energy levels previously observed.  

\section{Radiation hardness results}
\begin{figure}
  \begin{center}
    \mbox{\epsfig{file=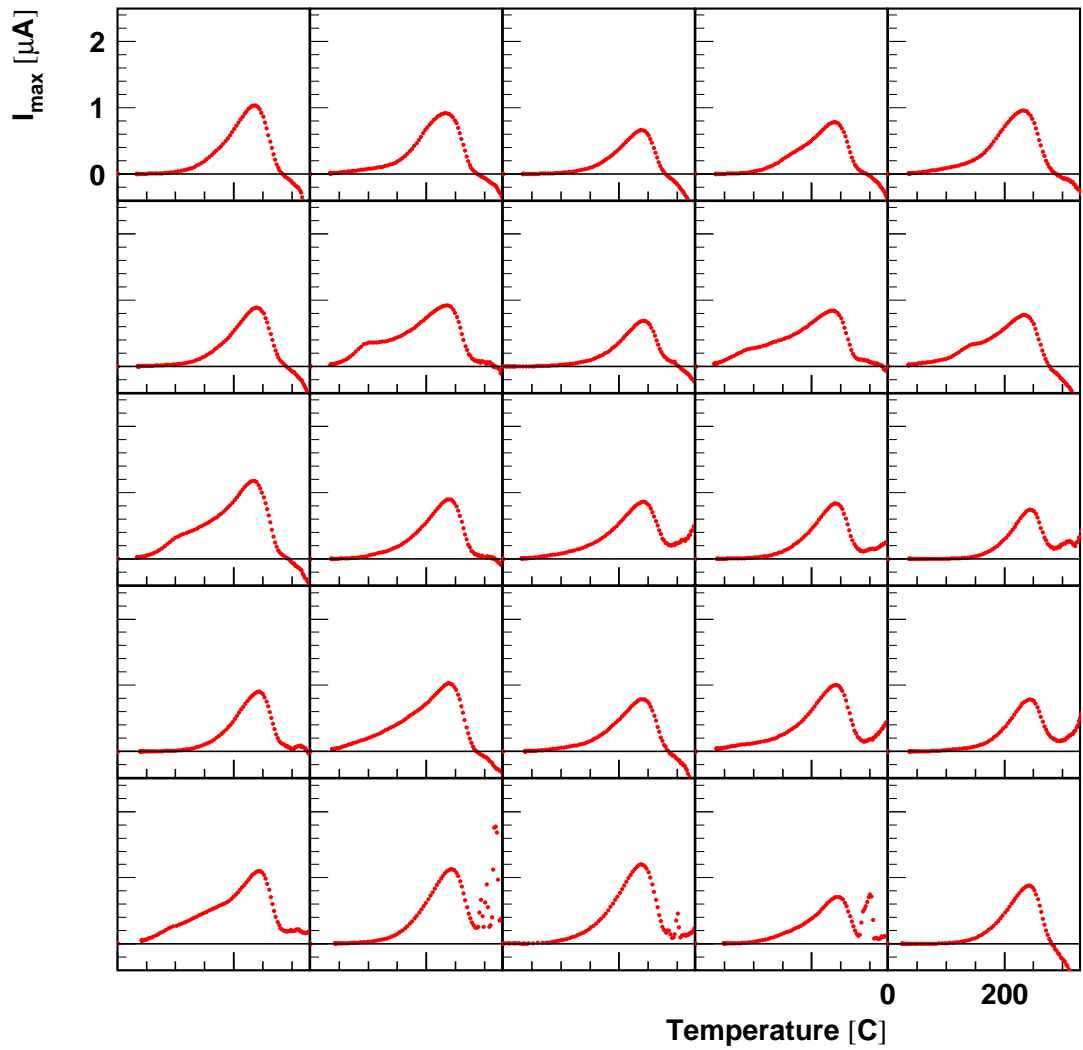,width=16cm}}
    \caption{\label{fig-scan}
   TSC curves for an area of $500\times 500 \mu{\rm m}^2$ scanned
   in steps of $100\times 100 \mu{\rm m}^2$.
      } 
  \end{center}
\end{figure}

A surface of approximately 1~mm$^2$ of diamond was scanned in
steps of $100\times 100 \mu{\rm m}^2$.  A TSC measurement was 
performed at each position.  The background subtraction and the photon
flux correction were applied to the measured signal.  Three scans were 
performed: before irradiation, after a dose of 0.2~MGy and after a
dose of 1.4~MGy.  Due to time constraints, a surface of only
$\sim$~0.5~mm$^2$ was scanned after the last dose.
The surface scanned was not the same for all scans.
Figure~\ref{fig-scan} shows an example of a TSC scan.  The figure
shows TSC curves after background subtraction.  These data were taken 
before irradiation.  One can see variations
in the background, leading to the 12\% systematic error on the 
peak measurement mentioned in Section~\ref{sect-signal}.  Some large
fluctuations are also seen at large temperatures, above the TSC peak.
These fluctuations do not cause problems in the analysis.
It is also interesting to note that in some cases, what appears to
be a second small peak at around 150~$^{\circ}$C is also observed.  This
could be due to the presence of an additional trap level.  Given the
size of these peaks and the fact that they do not appear at every 
scanned position, they were not used for testing radiation hardness.  
An example of such a peak has been discussed in Section~\ref{sect-levels}.

\begin{figure}
  \begin{center}
    \mbox{\epsfig{file=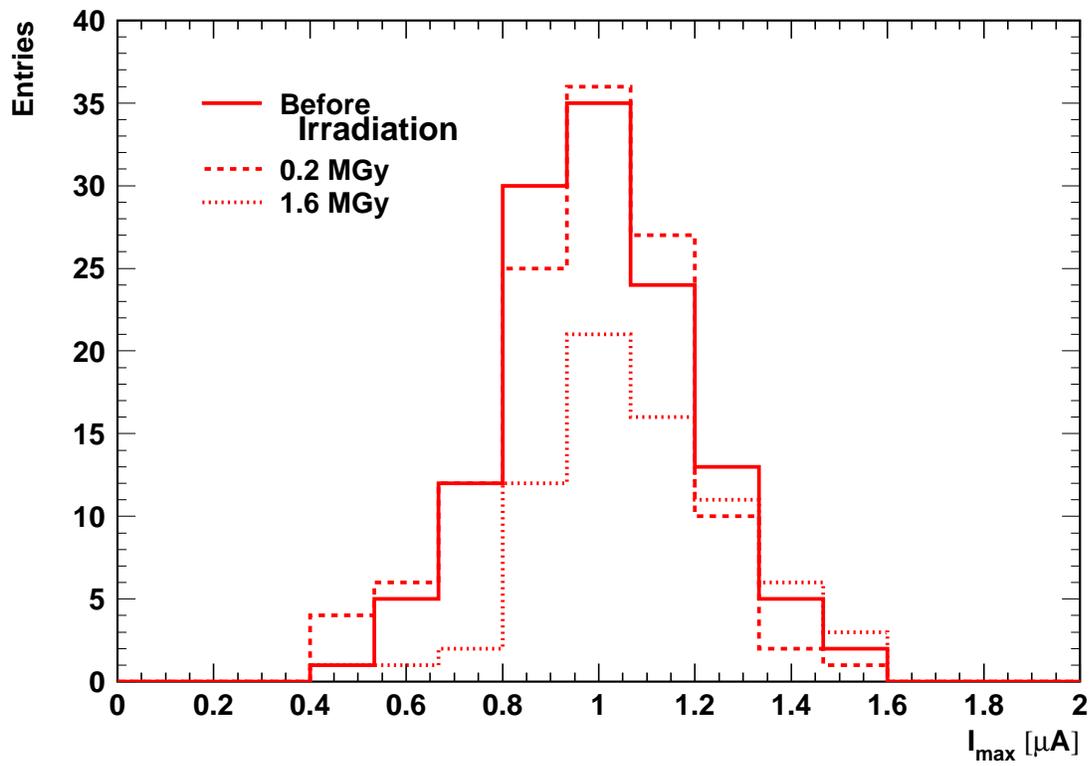,width=16cm}}
    \caption{\label{fig-result}
    Distribution of the current at the TSC peak for 
    several doses of electromagnetic radiation.
      } 
  \end{center}
\end{figure}

Figure~\ref{fig-result} shows the distribution of $I_{\rm max}$, the
current at the TSC peak, for each scan.  No effect on the TSC peak 
distribution as a function of the radiation dose was observed.
The RMS/mean values obtained
for the different doses are: 0.2/1.0, 0.2/0.97 and 0.21/1.1.  
The distribution of points is therefore a little wider than the
systematic error, although not significantly.  Since the CVD diamond
material is made of a large number of small crystals having 
several tens of microns in lateral size, the measured signal is
an average over several crystals.  For this reason, the measurement 
is not sensitive to the granularity of the material.

\begin{figure}
  \begin{center}
    \mbox{\epsfig{file=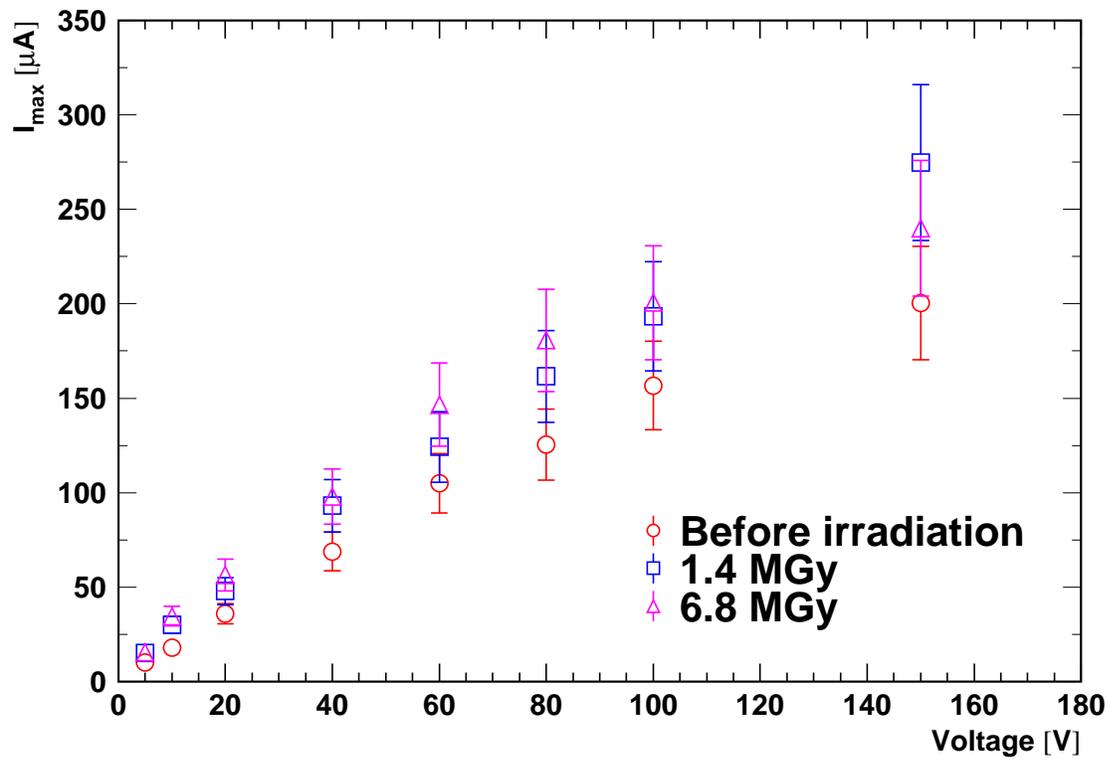,width=16cm}}
    \caption{\label{fig-rampv}
   Voltage dependence of the TSC peak for several
   doses of electromagnetic radiation.
      } 
  \end{center}
\end{figure}

The full surface of a second diamond (\#2) was irradiated in two
consecutive steps of 1.4~MGy and 5.4~MGy.  TSC measurements were performed
before irradiation and after each exposition to the beam.  TSC curves
were taken for different voltages.  The TSC peaks obtained were corrected for 
background and photon flux and are shown in Figure~\ref{fig-rampv}.
The rising behavior of the TSC peak as a function of the applied voltage 
corresponds to the increasing mobility for higher voltages.  A good agreement between 
data points taken after different doses of irradiation is observed.

\begin{figure}
  \begin{center}
    \mbox{\epsfig{file=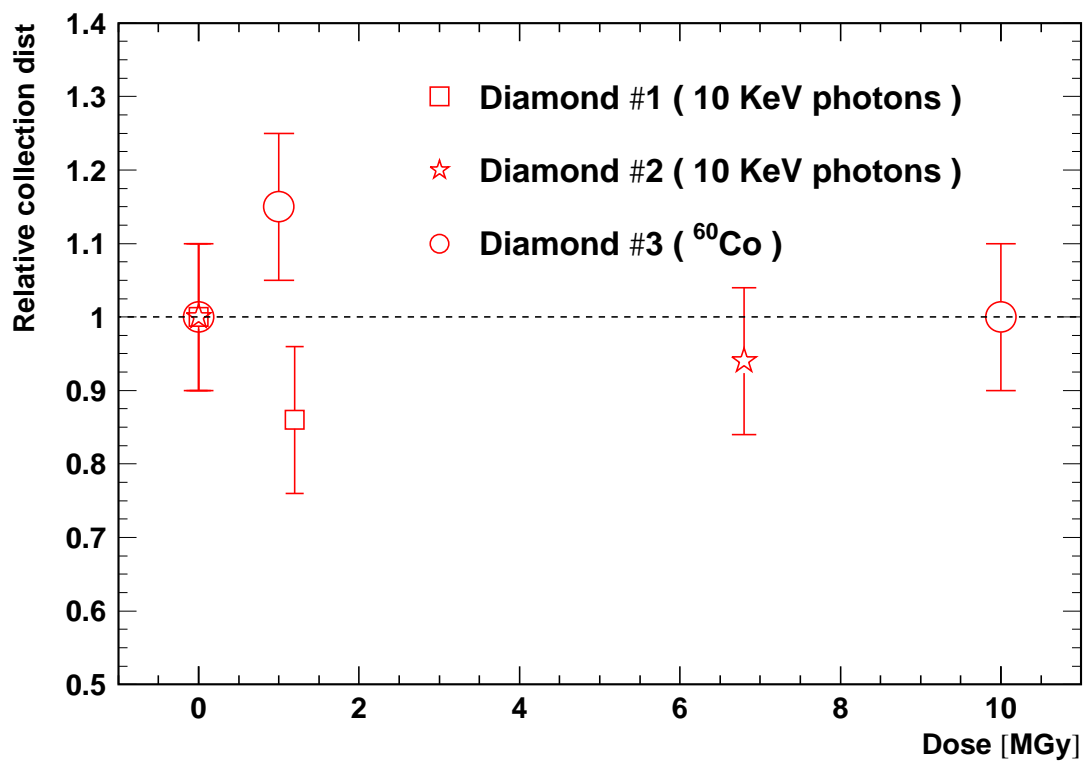,width=16cm}}
    \caption{\label{fig-colldose}
  The ratio of the collection distance measured after irradiation to the
  collection distance measured before irradiation
  for each diamond and each irradiation period.
      } 
  \end{center}
\end{figure}

Table~\ref{tab-colldist} shows the collection distance measurements
performed on the different diamond samples for various doses.
The ratio of the collection distance measured after irradiation to the
collection distance measured before irradiation is shown in 
Figure~\ref{fig-colldose} for each diamond and each irradiation period.
In the case of diamonds~\#1 and~\#2, which were used for the Hasylab
measurements, the collection distance was measured before the
TSC scans and after the last irradiation.  Diamond~\#3 was measured
before and after being sent to a $^{60}$Co irradiation facility.
The errors on the collection distances are mainly systematic and 
represent the reproducibility of the measurements.  They were
estimated at 10\% for each measurement, 
by comparing several measurements recorded at different times.
Several thousand data points, each corresponding to the passage of
a single minimum ionizing particle, were taken for each measurement,
so that the statistical error is negligible compared to the 
systematic error.  Although diamond~\#3 shows large variations
in its collection distance, no evidence for defects due to radiation
was observed.

\begin{table}[ht]
  \caption{\label{tab-colldist}
    Collection distance measurements of the different diamond
    samples after various doses of electromagnetic radiation.
    (*):~The first measurement of diamond~\#2 (before irradiation) was only 
    performed in an unprimed state, after only a few hours of irradiation 
    by a $^{90}$Sr source.  The same conditions were repeated for the
    second measurement (after 6.8~MGy).
    }
  \begin{center}
    \begin{tabular}{|l|l|l|l|}
      \hline
 Sample    & Dose (MGy) & $\gamma$ energy  & Coll. distance ($\mu$m) \\ \hline\hline
   \#1     &  0         & --               & $65 \pm 7$   \\ \hline
   \#1     &  1.6       & 10 keV           & $56 \pm 6$   \\ \hline\hline
   \#2 *   &  0         & --               & $31 \pm 3$   \\ \hline
   \#2     &  6.8       & 10 keV           & $29 \pm 3$   \\ \hline\hline
   \#3     &  0         & --               & $65 \pm 7$   \\ \hline
   \#3     &  1         & 1.17 \& 1.33 MeV & $75 \pm 8$   \\ \hline
   \#3     &  10        & 1.17 \& 1.33 MeV & $65 \pm 7$   \\ \hline
    \end{tabular}
   \end{center}
\end{table}

In summary, the collection distance measurements and the TSC measurements
show no indication of a degradation of the diamond quality as a function
of the radiation dose.  The collection distance remains constant within errors
up to at least 10~MGy and the TSC peak, which is indicative of the energy and 
the number of traps in the material, remains stable up to at least 6.8~MGy.

\section{Conclusions}
CVD diamond films were submitted to large electromagnetic radiation doses.  
A surface scan of thermally stimulated current measurements was performed
to test the stability of the trap level density after various irradiation doses.  
The collection distance of these diamond samples was also measured before
and after these irradiation periods.  The properties of the 
diamonds were found to remain unchanged after irradiation doses 
of up to 6.8~MGy of 10~keV photons and of up to 10~MGy of photons with
1.17~MeV and 1.33~MeV.
Such a resistance to electromagnetic radiation justifies the use of
diamond as material for future detectors such as the low-angle calorimeter
of TESLA.

\section*{Acknowledgements}
We would like to thank the staff of Hasylab for hosting us for the 
duration of our experiment.  A special thanks to C.~Paulmann who
was responsible for the experimental area and to G.~Kitis and P.~Gonon for 
discussions on the TSC analysis.  We would also like to thank A.~Oh for 
his help.


\newpage

\newpage

\end{document}